\documentclass[aps,prl,superscriptaddress,amsmath,amssymb,preprintnumbers,twocolumn,showpacs]{revtex4}
\usepackage{bm}
\usepackage{braket}
\usepackage{graphicx,color}
\newcommand{\NTT}{NTT Basic Research Laboratories, NTT Corporation, 3-1
Morinosato-Wakamiya, Atsugi, Kanagawa, 243-0198, Japan.}
\newcommand{\keio}{Graduate School of Fundamental Science and Technology, Keio University,Yokohama 223-8522, Japan}
\begin{document}


\title{Spin Amplification with Inhomogeneous Broadening}


\author{Suguru Endo}
\affiliation{\keio}
\author{Yuichiro Matsuzaki}
\affiliation{\NTT}
\author{William J. Munro}\affiliation{\NTT}
\author{Shiro Saito}\affiliation{\NTT}


\begin{abstract}
A long-lived qubit is usually well-isolated from all other systems and the environments, and so is not easy to couple with  measurement apparatus.  It is sometimes difficult to
   implement reliable projective measurements on such a qubit. One potential solution is spin amplification with many ancillary qubits. Here, we propose a spin amplification technique, where the ancillary qubits state change depending on the state of the target qubit. The technique works even in the presence of inhomogeneous broadening. We show that  fast and accurate
   amplification is possible even if the coupling and frequency of the ancillary qubits is inhomogeneous. Since our scheme is robust against realistic imperfections, this could provide a new mechanism for reading out a single spin that could not have been measured using the previous approaches.
\end{abstract}
\maketitle
Accurate quantum measurements and long coherence times are typical requirements
for quantum information processing. It is essential to perform quantum
measurements when we require quantum error correction\cite{correction} or need quantum
feedback technology \cite{feedback}.

Moreover, long coherence time is important in terms of retaining the quantum information during
the waiting time for the next operations. A lot of progress has been made on the technique for fabricating a long-lived qubit with many physical systems, and the coherence time has been significantly improved. For example, an electron spin has a coherence time
longer than a second\cite{A.M.Tyryshkin}, while the coherence time of a nuclear spin is
longer than a minute\cite{Kamyar2013}. Even for an artificial two-level system such as a superconducting qubit, the coherence time is of the order of tens of microseconds\cite{Rigetti}.

It is sometimes difficult to read out the signal from a long-lived qubit, because long-lived qubits  do not generally couple well with the measurement apparatus.
 To increase the coherence time, the qubit should be isolated so that it has a tiny coupling strength with other external systems. On the other hand, to read out a qubit with high fidelity requires  a strong coupling with the measurement apparatus,
 which may be contradicted by the fact that the qubit is isolated from other systems.

One solution for this problem is spin amplification. If we
couple the target qubit to read out with many ancillary qubits, we can
use the ancillary qubits to increase the effective signals, even if there is little 
individual coupling between the target qubit and ancillary qubits.  If the target qubit is in the ground state, the ancillary qubits remain in the
ground state, while the ancillary qubits will be excited if the
target qubit is in the excited state. There has been a lot of researches in this direction.
A simple quantum circuit, which consists of controlled-NOT gates, is a
well-known  example \cite{ganso,ghz,entanglement} . However, in this scheme each spin must be  independently accessible and interact with the target qubit and it is difficult to increase the scale of such a scheme. Although there have been some proposals as regards a scalable spin amplification, they require a special
configuration for ancillary qubits with tailored couplings\cite{C.A,A.Kay,J.A,J.S}
and/or the protocol is fragile with respect to certain some imperfections such as the
inhomogeneous broadening or thermal excitation of the ancillary qubits\cite{Close2011,Negoro2011}.  

 In this paper, we suggest another way to amplify the signal of the
target qubit where many ancillary qubits are collectively coupled with the
target qubit. Since this is a simple experimental setup, there are many
physical systems that have this configuration.
A superconducting qubit coupled with a spin ensemble is one 
candidate for realizing our scheme
\cite{Saitosan,zhu,Kubosanprl,zhusan}. A nuclear (or electron) spin coupled with many
satellite nuclear (or electron) spins with different frequencies is another candidate \cite{Weimer}. 

In this paper, we address the problem of the inhomogeneous
broadening in particular\cite{inhomo1,inhomo2}. In reality,
it is very difficult to fabricate identical qubits so that every ancillary qubit can have a different frequency and also have a different coupling strength with the target qubit.
Interestingly, in our scheme, 
we can realize rapid and reliable spin amplification even under the
influence of inhomogeneous broadening. The target qubit is coupled
with the collective mode of the ancillary qubits.
On the other hand, the collective mode of the ancillary qubits is
coupled with the subradiant modes of the qubits, which induces 
dissipation.
If we try to realize coherence transfer between the target spin and the
collective mode,
the existence of the subradiant modes induces decoherence, which means the fidelity of the coherent transfer decreases. 
However, with spin amplification, coherence is not required
because the purpose is to read out the target qubit,
which eliminates the diagonal term of the density matrix of the target spin
if the quantum measurement is successfully implemented.
Moreover, we have shown that coupling between the collective mode
and subradiant mode only induces an energy transfer from the former to the latter that can be also measured as well as the
former.
Our technique can provide a reliable readout of a qubit coupled with
many inhomogeneous ancillary qubits.

The Hamiltonian of our hybrid system composed of the target qubit and
ancillary qubits can be described by
\begin{figure}[h!]
\centering
\includegraphics[width=5cm]{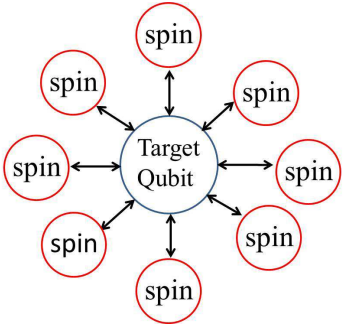}
\caption{Schematic of our hybrid system. The central target qubit is
 collectively coupled with the ensemble of ancillary qubits. The
 target qubit can be readout via the ancillary qubits after the
 amprification technique}
\label{tatai}
\end{figure}
\begin{align}
\hat{H}_{\text{TA}}=&\frac{\omega_T}{2}\hat{\sigma}_z+\sum_{j=1}^{N}g_j(\hat{\sigma}^-\hat{\sigma}_j^++\hat{\sigma}^+\hat{\sigma}_j^-)+\sum_{j=1}^{N}\frac{\omega_{j}}{2}\hat{\sigma}_{z j} \\
&+\lambda_d\mathrm{cos}(\omega_d t)(\hat{\sigma}^++\hat{\sigma}^-)  \nonumber
\label{hamiltonian}
\end{align}
where we set $\hbar=1$. Here, $\hat{\sigma}_{z}$ denotes the Pauli z operator for the target qubit and
$\hat{\sigma}^{-(+)}$denotes the ladder operator. $\omega_T$ and $\omega_j$ denote the resonant frequencies of the target qubit and the $j$ th spin, respectively. $\hat{\sigma}_j^{-(+)}$ and $\hat{\sigma}_{zj}$ denote the ladder operator and Pauli z operator
of the $j$ th ancillary qubit, respectively. It is worth mentioning
that, as long as most of the ancillary qubits are in the ground state
and only a small portion of the ancillary qubits are excited for the amplification,
we can consider the ancillary qubit to be a harmonic oscillator. So we can replace $\hat{\sigma}^{-(+)}_j$ with the annihilation(creation) operator  $\hat{a}_j^{(\dag)}$.
$g_j$ is a real number and denotes the coupling strength
between the target qubit and
the $j$ th ancillary qubit. $\lambda _d$ denotes the Rabi
frequency of the microwave applied to the target qubit.
We assume that the coupling strength $g_j$ is much smaller than the
inhomogeneous broadening width of this system.
By employing a rotating frame defined by
$\hat{U}(t)=\mathrm{exp}[-i\omega_dt(\frac{\hat{\sigma}_z}{2}+\sum_{j=1}^N \hat{a}_j^\dag\hat{a}_j)]$
and performing a rotating wave approximation, we have
\begin{align}
\hat{H}_{RWA}\approx&\frac{\omega_{T}-\omega_d}{2}\hat{\sigma}_z+\sum_{j=1}^{N}g_j(\hat{\sigma}^-\hat{a}_j^\dag+\hat{\sigma}^+\hat{a}_j)  \\ 
&+\sum_{j=1}^{N}(\omega_j-\omega_d)\hat{a}_j^\dag \hat{a}_j  
+\frac{\lambda_d}{2}(\hat{\sigma}^++\hat{\sigma}^-)  \nonumber
\end{align}

We will now model the effect of inhomogeneous
broadening with a Lorentzian distribution of the resonant frequencies of the spin ensemble as
\begin{align}
f(\omega)=\frac{1}{\pi}\frac{\gamma/2}{(\omega-\bar{\omega})^2+(\gamma /2)^2}
\label{loren}
\end{align}   
where, $\gamma$ denotes the width of the Lorentzian, and
$\bar{\omega}$ denotes
the average spin ensemble resonant frequency. We define the collective mode as follows
\begin{align}
\hat{A}\equiv \frac{1}{G}\sum_{j=1}^{N}g_j \hat{a}_j
\label{collective}
\end{align}
where $G=\sqrt{\sum_{j=1}^{N} g_j^2}$. Since we are mainly interested in the target qubit and
collective mode, we can trace out the sub-radiant modes. In this case, the collective Hamiltonian can
be written as \cite{Dinitz}
\begin{align}
 \hat{H}&=\hat{H}_c +\hat{H}_d\ \ \ \ \ \ \ \ \ \ \ \ \ \ \ \ \ \ \ \ \ \ \ \ \ \ \ \ \ \ \label{hcc}
  \\
\hat{H}_c&=\frac{\omega_{T}-\omega_d}{2}\hat{\sigma}_z+G(\hat{A}\hat{\sigma}^{+}+\hat{A}^\dag
 \hat{\sigma}^-)+(\bar{\omega}-\omega_d)\hat{A}^\dag \hat{A}\label{hc} \\
 \hat{H}_d&=\frac{\lambda_d}{2}(\hat{\sigma}^++\hat{\sigma}^-) \ \ \ \ \ \ \ \ \ \ \ \ \ \ \ \ \ \ \ \ \ \ \ \ \ \ \ \ \ 
\end{align}
The width of the Lorentzian $\gamma$ can be incorporated in the dynamics as the decay term of the master equation, which we will discuss later.

In our scheme, we will drive the target qubit with a microwave pulse
for the spin amplification. The collective mode coupled with the
target qubit may differ from the collective mode coupled with
the microwave line, which is called a mode mismatch\cite{mismatch}. This problem means that driving the collective mode by the coupling between the microwave line and ancillary qubits is not straightforward. On
the other hand, as we explain later, it is
possible to drive the collective mode of the ancillary qubits by
driving the target qubit.
We have developed an efficient technique for calculating the number of 
excitations in the spin ensemble after driving. If we include the all
freedom of N ancillary qubits, we need to deal with 
$2^{N+1}$ sized matrix, which is intractable. On the other hand, with the technique
that we
propose here, we need to solve a Hamiltonian that includes a single
qubit and a harmonic oscillator, which is much easier to simulate.
Using Eq.\ref{hamiltonian} as a basis, we can write the Heisenberg equation,
\begin{align}
\frac{d}{dt}\hat{a}(t)_j=-ig_j\hat{\sigma}(t)^- -i(\omega_j-\omega_d-i\frac{\Gamma}{2})\hat{a}(t)_j
\label{heisenberg}
\end{align}
where we add the energy decay term $\Gamma $ of the ancillary qubit.
However,  this decay rate is usually negligible
compared with that of other
noise.
For example, the energy relaxation time of an electron
such as the NV$^-$ center exceeds
200 seconds \cite{inhomo4} . So we adopt a limit of $\Gamma \rightarrow 0$, later.  
Now
\begin{align}
\frac{d}{dt}\sum_{j=1}^N \hat{a}(t)_j^\dag
 \hat{a}_j=iG(\hat{\sigma}^+\hat{A}(t)
 -\hat{A}(t)^\dag \hat{\sigma}^-)-\Gamma \sum_{j=1}^N \hat{a}(t)_j^\dag \hat{a}(t)_j
\label{col1}
\end{align}
Now, taking the Laplace transforms of both sides of 
Eq.\ref{heisenberg}, under the initial conditions
$a_j(0)=0,\sigma^-(0)=0$, we have
\begin{align}
\hat{\alpha}_j(s)=-\frac{ig_j\hat{\beta}(s)}{s+i(\omega_j-\omega_d-i\frac{\Gamma_s}{2})}
\label{laplace}
\end{align}
Here, $\hat{\alpha}_j(s)=\mathcal{L}[\hat{a}_j(t)],\hat{\beta}(s)=\mathcal{L}[\hat{\sigma}^-(t)].$
Using the Lorentzian distribution  (\ref{loren}) and adopting residual integration, we obtain
\begin{align}
\frac{1}{G}\sum_{j=1}^N g_j\hat{\alpha}_j(s)=-\frac{i G \hat{\beta}(s)}{s+i(\bar{\omega}-\omega_d-i\frac{\gamma+\Gamma}{2})}
\label{residue}
\end{align}
Taking the inverse Laplace transformation of both sides, we obtain
\begin{align}
iG\sigma^-(t)=-\frac{d}{dt}\hat{A}(t)-i(\overline{\omega}-\omega_d-i\frac{\gamma+\Gamma_s}{2})\hat{A}(t)
\label{col2}
\end{align}
By using Eq.\ref{col2} and Eq.\ref{col1},  in the limit of $\gamma>>\Gamma$ we obtain
\begin{align}
\frac{d}{dt}(-\hat{A}^\dag(t) \hat{A}(t) +\sum_{j=1}^N \hat{a}_j^\dag(t) \hat{a}(t)_j)=\gamma \hat{A}^\dag(t) \hat{A}(t)
\label{col3}
\end{align}
This means that the rate of increase of the subradiant modes ($-\hat{A}^\dag(t) \hat{A}(t)
+\sum_{j=1}^N \hat{a}_j^\dag(t) \hat{a}_j(t)$) depends on the collective mode
($\hat{A}^{\dagger }(t)\hat{A}(t)$) and the inhomogeneous width ($\gamma $).
By integrating Eq.\ref{col3}, we obtain
\begin{align}
\sum_{j=1}^N \hat{a}(t)_j^\dag \hat{a}(t)_j=\hat{A}(t)^\dag \hat{A}(t)+\gamma \int_0^t\hat{A}(t^\prime)^\dag \hat{A}(t^\prime)dt^\prime
\label{col4} 
\end{align}
By calculating he dynamics of A(t), we can estimate the excitation number $\sum_{j=1}^N\hat{a}_j(t)^\dag \hat{a}_j(t)$



We intuitively explain how our amplification works.
If the detuning between the energy of the target qubit and the frequency
of the collective mode is much larger than the coupling strength, we can
derive the dispersive Hamiltonian as follows \cite{intro} . 
\begin{align}
\hat{H}_{disp}=\frac{\omega_{T}-\omega_d}{2}\hat{\sigma}_z+\frac{G^2}{\Delta}\hat{\sigma}^+\hat{\sigma}^-
+
 (\bar{\omega}-\omega_d+\frac{G^2}{\Delta}\hat{\sigma}_z)\hat{A}^\dag
 \hat{A}
\label{dispersive}
\end{align}
where $\lambda _d=0$ and $\Delta=(\omega_{T}-\omega_d)-(\bar{\omega}-\omega_d)$.
This means that, depending on the state of the target qubit, there is a
dispersive energy shift as follows.
 \begin{equation}
 \bar{\omega}  \Rightarrow  \bar{\omega}  \pm \frac{G^2}{\Delta}
 \label{disp}
 \end{equation}
So, if the driving frequency is $\bar{\omega} +\frac{G^2}{\Delta}$ and
the target qubit is in an excited state, the ensemble of the
ancillary qubits is efficiently excited.
On the other hand, if the driving frequency is $\bar{\omega}
+\frac{G^2}{\Delta}$ and the target qubit is in the ground state,
the spin ensemble is not significantly affected by the driving caused by
the detuning. Throughout this paper,
we fix the driving frequency as  $\bar{\omega} +\frac{G^2}{\Delta}$.

Although we later  solve the master equation
numerically, it is also possible to describe the dynamics of the collective
mode in an analytical way by using an approximation.
By diagonalizing the $H_c$ in Eq.\ref{hc}, we obtain\cite{intro}
\begin{align}
&\hat{H}_c=\sum_n  \hat{H}_n \\
&\hat{H}_n=\Omega_+ (n)\ket{+,n} \bra{+,n}+\Omega_-(n)\ket{-,n} \bra{-,n} \\
&\Omega_{\pm}(n)=(n+\frac{1}{2})(\bar{\omega}-\omega_d)\pm\frac{\sqrt{\Delta^2+4G^2(n+1)}}{2} \\ 
\end{align}
  The eigenstate of the Hamiltonian is 
\begin{align}
&\ket{+,n}=\mathrm{cos}(\phi_n/2)\ket{e,n}+\mathrm{sin}(\phi_n/2)\ket{g,n+1} \\
&\ket{-,n}=-\mathrm{sin}(\phi_n/2)\ket{e,n}+\mathrm{cos}(\phi_n/2)\ket{g,n+1} \\
&\phi_n=\mathrm{tan}^{-1}\bigg(\frac{2G\sqrt{n+1}}{\Delta}\bigg)
\label{eigenstate}
\end{align}
where $|e\rangle $ ($|g\rangle $) denotes the excited
(ground) state of the target qubit and $|n\rangle $ denotes the number
state of the collective mode.
Since we can diagonalize the Hamiltonian $H_c$,
we can efficiently use an interaction picture where we define
$\hat{U}=e^{-iH_ct}$ and $H_I(t)=\hat{U}^{\dagger
}\hat{H}_d\hat{U}$.
By using the condtion $\Delta \gg G$,
we obtain
\begin{align}
 H_I(t)\simeq \lambda_{eff} \Big{(} |e\rangle \langle
  e|(\hat{A}+\hat{A^{\dagger }})  
-|g\rangle \langle
  g|(\hat{A}e^{i\frac{2G^2}{\Delta }t} 
+\hat{A^{\dagger
  }}e^{-i\frac{2G^2}{\Delta }t} )\Big{)}
\end{align}
where $\lambda_{eff}= \frac{\lambda_d}{2}\frac{G}{\Delta}$ denotes
the effective Rabi frequency of the collective mode.
If the initial state of the target qubit is $|e\rangle $ 
and it remains unchanged during the time
evolution, we can derive the Hamiltonian for the collective mode
 $ H^{(\text{anc})}_I(t)\simeq \lambda_{eff} 
  (\hat{A}+\hat{A^{\dagger }}) $.
  On the other hand, if the target qubit is kept in
  the ground state, we obtain
  $ H_I(t)\simeq -\lambda_{eff}
  (\hat{A}e^{i\frac{2G^2}{\Delta }t}+\hat{A^{\dagger
  }}e^{-i\frac{2G^2}{\Delta }t} )$.
By returning to the Schrodinger picture, we obtain
\begin{align}
\hat{H}_{anc}= (\bar{\omega}-\omega_d\pm \frac{G^2}{\Delta})\hat{A}^\dag
 \hat{A}\pm \lambda_{eff}(\hat{A}+\hat{A}^\dag)
\label{hamilancilla}
\end{align}
where the sign $\pm$ depends on the state of the target qubit.
With this Hamiltonian, we obtain the Heisenberg equation of $\hat{A}$ as follows,
\begin{align}
\frac{d}{dt}\hat{A}=-i(\bar{\omega}-\omega_d\pm \frac{G^2}{\Delta}-i\frac{\gamma}{2})\hat{A}-i\lambda_{eff}
\end{align}
where we added the energy relaxation term $-i \frac{\gamma}{2} \hat{A}$.
When the target qubit is in an
excited state, we solve this equation
analytically with $\omega_d=\bar{\omega}+\frac{G^2}{\Delta}$, and obtain
\begin{align}
\braket{\hat{A}^\dag(t)\hat{A}(t)}=\frac{4\lambda_{eff}^2}{\gamma^2}\bigg(1-e^{-\frac{\gamma}{2} t})^2
\end{align}
On the other hand, when the target spin is in the ground state, we
obtain 
\begin{align}
\braket{\hat{A}^\dag(t)\hat{A}(t)}=\frac{\lambda_{eff}^2}
 {(\frac{2G^2}{\Delta})^2+(\frac{\gamma}{2})^2} \bigg[1-2\mathrm{cos}\bigg(\frac{2G^2}{\Delta}t\bigg) e^{-\frac{\gamma}{2}t}+e^{-\gamma t}\bigg]
\label{targetup}
\end{align}
It should be noted that, if the effective Rabi frequency is far
smaller than the dispersive energy shift such as $\lambda _{eff}\ll
\frac{G^2}{\Delta }$, the excitation of the collective mode is
significantly suppressed when the target qubit is prepared in the ground state.
So we retain this condition 
throughout this paper.

\begin{figure}[h!]
\centering
\includegraphics[width=8.8cm]{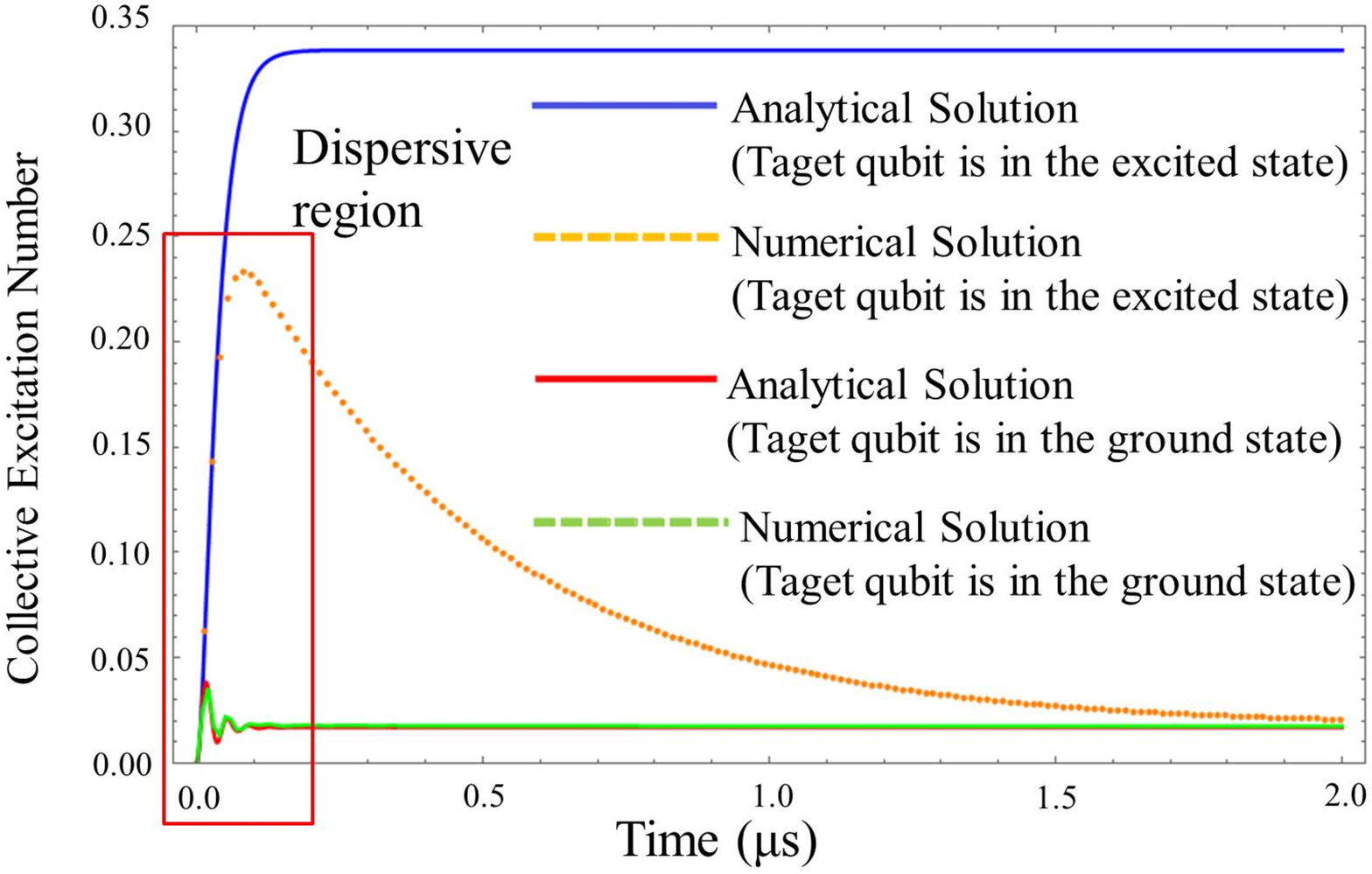}
\caption{Collective mode excitation number
 against time.
 Solid lines show analytical solutions and dashed lines show numerical calculations where we set $\gamma/2\pi =12.5$  MHz, $G/2\pi
 $=75  MHz, $\Delta/2\pi =412.5$ MHz, and $\lambda_d/2\pi =40$  MHz. These are typical parameters of the hybrid system composed of superconducting flux qubit and NV- centers\cite{zhu}. When obtaining
 analytical solutions, we use a dispersive Hamiltonian where the state
 of the target qubit is not changed, which is a valid approximation for
 a short time region (dispersive region). However, for a numerical
 simulation, the excited state of the target qubit decays into the ground
 state due to the flip flop term of the JC Hamiltonian and subradient mode,
 which leads a difference from the analytical approach for a longer time
 scale.}
\label{collective2}
\end{figure}

\begin{figure}[h!]
\centering
\includegraphics[width=8.6cm]{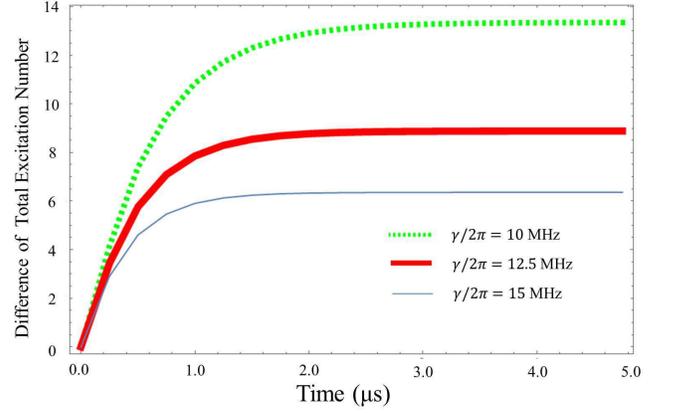}
\caption{Spin amplification process in our scheme.
 Depending on the initial state of the target
 qubit ($|e\rangle$ or $|g\rangle $), the ancillary qubit excitations numbers differ, and
 we plot this difference against time. 
 We set the parameters at $G/2\pi =75$  MHz, $\lambda_d/2\pi =40$  MHz, $\Delta/2\pi =412.5$ MHz
 }
\label{individualmode}
\end{figure}
We calculate the time evolution of the collective mode by solving the following master equation.
\begin{eqnarray}
 \frac{d \hat{\rho} }{dt}=-i[\hat{H},\hat{\rho} ]
  +\frac{\gamma}{2}[2\hat{A}\hat{\rho}\hat{A}^\dag-\hat{A}^\dag\hat{A}\hat{\rho}-\hat{\rho}\hat{A}^\dag\hat{A}]\label{lindblad}
\end{eqnarray}
It is worth mentioning that the inhomogeneous broadening of the frequency can be
considered an energy loss of the collective mode \cite{Dinitz} .
In Fig.\ref{collective2},
we plot the time evolution of the collective mode by using the
analytical approach described above and a numerical
approach to solve the master equation of Eq. \ref{lindblad}.
We assume that the initial state is $|e,0\rangle $ or $|g,0\rangle $. In the analytical approach,
the state of the target spin does not change during the time evolution, because we use a
dispersive Hamiltonian.
When the initial state of the target qubit is an excited
state, the collective mode of the spin ensemble is excited efficiently
in both the numerical simulation and the analytical calculation for
a short time scale. However, since the energy of the
target qubit is gradually transfered to the subradiant modes of the
ancillary qubits, the target qubit decays into the ground state and so 
the collective mode will not be efficiently excited over a long time period in the numerical simulation \cite{enhanced} . This induces a disagreement between the analytical solution and numerical simulation in this regime.
On the other hand, when the initial state of the target qubit is the
ground state, the state of the target will not be significantly changed,
and so there is an excellent agreement between all $t$ values of the analytical solution and
numerical simulation where the excitation of the collective modes is
efficiently suppressed.

We plot the difference in the excitation population of the ensemble when
 the target qubit is in its excited and ground states
 in Fig.\ref{individualmode}.
Although the inhomogeneous broadening reduces the
efficiency of the amplification, we can still obtain a difference of more than 10
excitations in the ancillary qubits for $\gamma/2\pi  = 10$
MHz. This means that we will have a 10 times better signal with our
amplification technique than without any amplification, which
would provide us with a practical tool with which to read out a single qubit such as an
electron spin or a nuclear spin.

In conclusion, we have proposed a spin amplification technique that can be used in the presence of the inhomogeneous broadening. We considered a simple experimental system where a target qubit is collectively coupled with many ancillary qubits. We showed that, even if the ancillary qubits are inhomogeneous, we can amplify the signal from the target spin, and obtain a gain of 10 in such a system by applying a continuous microwave. Our technique opens the way to the fast and robust readout of a single spin.

\vspace{4mm}
We thank H. Yamaguchi and J. Hayase for valuable discussions. We thank T. Koike for encouraging S. Endo's research. This work was supported by Commissioned Research No. 158286  of National Institute of Information and Communication 287 Technology (NICT).

\end{document}